\documentclass[aps,prb,twocolumn,superscriptaddress]{revtex4-1}
\usepackage{amsmath,amsthm,amssymb,bm,color,comment}
\usepackage{graphicx}
\graphicspath{{./fig/}}
\begin{document}

\title{Combined experimental and theoretical studies on glasslike transitions in the frustrated molecular conductors $\theta$-(BEDT-TTF)$_2MM'$(SCN)$_4$}

\author{Yohei Saito}
\affiliation{Institute of Physics, Goethe University Frankfurt, 60438 Frankfurt (M), Germany}

\author{Owen Ganter}
\affiliation{Department of Physics and Center for Functional Materials, Wake Forest University, Winston-Salem, North Carolina 27109, USA}

\author{Chao Shang}
\affiliation{Department of Physics and Center for Functional Materials, Wake Forest University, Winston-Salem, North Carolina 27109, USA}

\author{Kenichiro Hashimoto}
\affiliation{Department of Advanced Materials Science, University of Tokyo, 277-8561 Chiba, Japan}
\affiliation{Institute for Materials Research, Tohoku University, 980-8577 Sendai, Japan}

\author{Takahiko Sasaki}
\affiliation{Institute for Materials Research, Tohoku University, 980-8577 Sendai, Japan}

\author{Stephen M. Winter}
\affiliation{Department of Physics and Center for Functional Materials, Wake Forest University, Winston-Salem, North Carolina 27109, USA}

\author{Jens M\"{u}ller}
\affiliation{Institute of Physics, Goethe University Frankfurt, 60438 Frankfurt (M), Germany}

\author{Michael Lang}
\affiliation{Institute of Physics, Goethe University Frankfurt, 60438 Frankfurt (M), Germany}

\begin{abstract}
We present results of the coefficient of thermal expansion for the frustrated quasi-two-dimensional molecular conductor $\theta$-(BEDT-TTF)$_2$RbZn(SCN)$_4$ for temperatures 1.5\,K $\leq T \leq$ 290\,K. A pronounced first-order phase transition anomaly is observed at the combined charge-order/structural transition at 215\,K. Furthermore, clear evidence is found for two separate glasslike transitions at $T_{\mathrm{g}}$ = 90--100\,K and $T_{\mathrm{g}}^\dag$ = 120--130\,K, similar to previous findings for $\theta$-(BEDT-TTF)$_2$CsZn(SCN)$_4$ and $\theta$-(BEDT-TTF)$_2$CsCo(SCN)$_4$, reported in T. Thomas \textit{et al}., Phys. Rev. \textbf{B} 105, L041114 (2022), both of which lack the charge-order/structural transition. Our findings indicate that these glasslike transitions are common features for the $\theta$-(BEDT-TTF)$_2MM^\prime$(SCN)$_4$ family with $M$ = (Rb, Cs) and $M^\prime$ = (Co, Zn), irrespective of the presence or absence of charge order. These results are consistent with our model calculations on the glasslike dynamics associated with the flexible ethylene endgroups of the BEDT-TTF molecules for various $\theta$-(BEDT-TTF)$_2MM^\prime$(SCN)$_4$ salts, predicting two different conformational glass transitions. Moreover, calculations of the hopping integrals show a substantial degree of dependence on the endgroups' conformation, suggesting a significant coupling to the electronic degrees of freedom. Our findings support the possibility that the glassy freezing of the ethylene endgroups could drive or enhance glassy charge dynamics.

\end{abstract}

\maketitle

\section{INTRODUCTION}
\begin{figure}[t]
  \includegraphics[width=8.6cm]{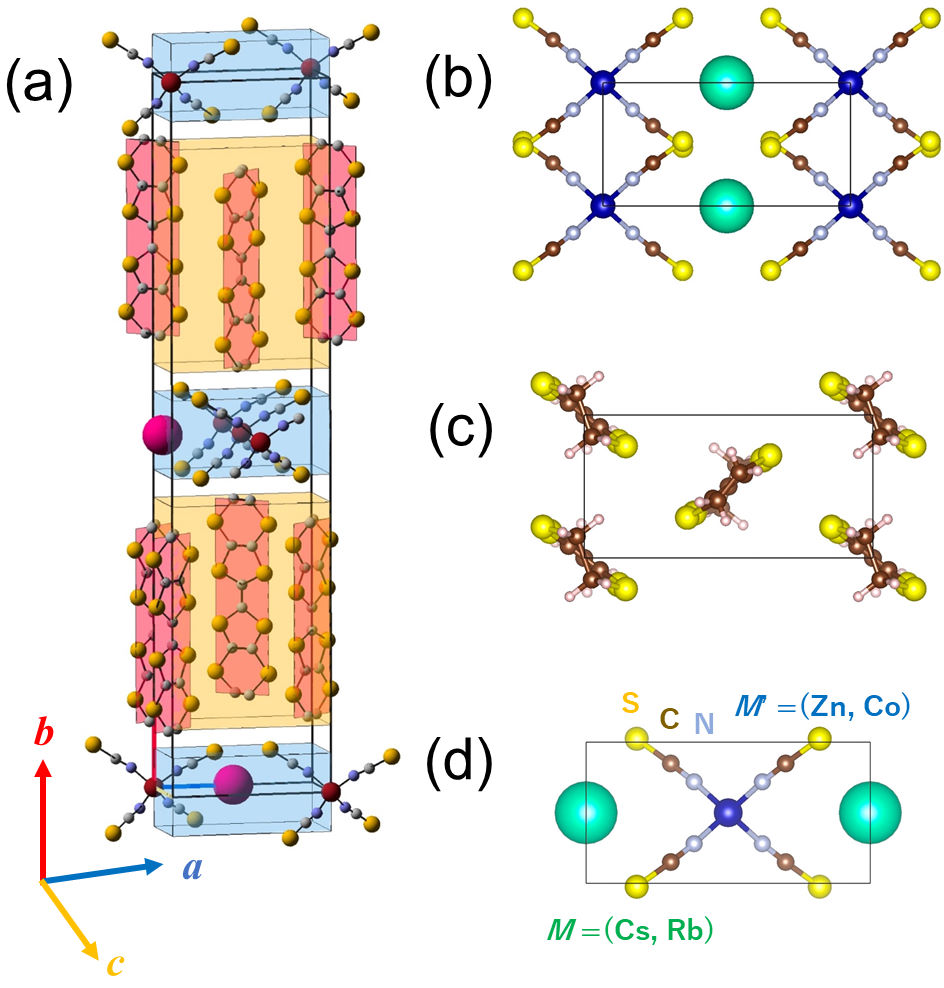}
  \caption{Crystal structure of $\theta$-(BEDT-TTF)$_2MM^\prime$(SCN)$_4$. (a) $MM^\prime$ = RbZn, RbCo, CsZn, and CsCo. Cross sections of the cation layer and anion layer (b)--(d): the BEDT-TTF in the middle of the unit cell is situated at the same position as the $M^\prime$(SCN)$_4$ complex in the bottom anion layer. The surrounding four BEDT-TTF molecules located at the corners of the unit cell do not have any corresponding atoms in the anion layer at the same positions. Instead, Cs ions are located in the anion layer between the BEDT-TTF positions halfway between the left and right sides of the unit cell. These diagrams illustrate the interface between the cation and anion layer. On the opposite side of the BEDT-TTF molecules the top anion layer is translated by [±1/2, ±1/2, 0] of a unit cell. Thus, the BEDT-TTF in the middle of the unit cell has Cs ions located above and below it halfway on the top and bottom sides of the unit cell. The four BEDT-TTF molecules on the corners of the unit cell have corresponding $M^\prime$(SCN)$_4$ complexes also centered on the corners of the unit cell. This arrangement suggests that the ethylene endgroups (EEGs) located in proximity to the Zn/Co complexes are more hindered than the EEGs near the monatomic ions.}
\label{Crystal_structure}
\end{figure}
Structural degrees of freedom are important factors for the emergence of various intriguing phenomena in condensed matter. In the case of the quasi-two-dimensional organic charge-transfer salts of the (BEDT-TTF)$_2X$ family, where BEDT-TTF (ET in short) stands for bis(ethylenedithio)tetrathiafulvalene and $X$ for a monovalent anion, a system of correlated $\pi$ electrons (holes) is formed, which can interact with the underlying lattice in several ways. First and foremost, this interaction involves the degree of frustration, dictated by the choice of the anion $X$, as well as a significant electron-phonon coupling, manifesting itself, e.g., in pronounced effects in the coefficient of thermal expansion at various electronic phase transitions \cite{Muller2002, DeSouza2008, Gati2016, Gati2018}. Besides the intra- and intermolecular vibrational modes, another structural degree of freedom is introduced by the flexible ethylene [C$_2$H$_4$] endgroups (EEGs hereafter) of the ET molecules as they can adopt two different conformations: when viewed along the central C=C bond of the ET molecules, their orientation is either eclipsed (E) or staggered (S). In many salts the population of the E and S states is thermally disordered at room temperature and the system tends to adopt one of the two possible conformations upon cooling \cite{Hartmann2014}. In some cases, however, the ordering of the EEGs cannot be completed for kinetic reasons, giving rise to a glasslike transition at a temperature $T_{\mathrm{g}}$ below which a short-range structural order becomes frozen in \cite{Muller2015}. Thus the conformational degree of freedom associated with the EEGs also includes the aspect of disorder in terms of a random lattice potential.

Since the mechanism of glass formation, which can occur by many different routes \cite{Angell1995,Debenedetti2001}, is of fundamental interest, the molecular conductors offer an exciting avenue for studying glassy effects in the presence of strongly correlated electrons and frustration.
In this regard, the layered charge-transfer salts $\theta$-(ET)$_2MM^\prime$(SCN)$_4$, with $M$ = (Rb,Cs,Tl) and $M^\prime$ = (Co,Zn), constitute a particularly interesting class of materials. In this $\theta$ polymorph the ET molecules form a 1/4-filled correlated electron system on a triangular lattice (see Fig.\,\ref{Crystal_structure}) with different degree of (charge) frustration, depending on the choice of $M$ and $M^\prime$\, \cite{Kagawa2017,Sasaki2017}. In these systems intriguing phenomena in the charge sector have been reported including charge order (CO) accompanied by a structural transition around $T_{\rm CO} = 200$\,K in the $MM^\prime$ = RbZn salt (denoted $\theta$-RbZn) \cite{Mori1998, Watanabe2004,Alemany2015,Hashimoto2022}. Interestingly, for the more strongly frustrated $\theta$-CsCo and $\theta$-CsZn, as well as for rapidly cooled $\theta$-RbZn, no CO transition is observed. Instead, glasslike charge dynamics was revealed in a variety of experimental observables which has been assigned to a charge-glass (CG) state \cite{Kagawa2013,Sato2014}. It was found that this newly-discovered electron glass and the dynamics of strongly-correlated electrons residing on a geometrically-frustrated lattice is surprisingly similar to a structural glass and relaxation in conventional glass-forming liquids \cite{Sasaki2017,Sato2017,Murase2022}. The convenient time and temperature scales of these materials and the possibility of inferring volume fractions from easily accessible charge transport (resistance measurements) alone \cite{Sasaki2017} were suggested to enable to investigate very fundamental problems in glass physics like aging, memory effects, cooperativity, or the presence or absence of an underlying true phase transition from a different perspective.
Important open problems that are currently discussed are a classical to quantum crossover in the electron glass \cite{Murase2022,Fratini2023} depending on the degree of frustration and --- important for this work --- the interrelation between structural and electronic glassiness \cite{Thomas2022}.

In these previous works the glassy charge dynamics was discussed in purely electronic terms, where the key parameter that determines the glass-forming ability is the degree of frustration of the charges residing on the underlying triangular lattice \cite{Kagawa2017,Sasaki2017}, the latter being determined by the choice of the atoms $MM^\prime$ in the anion layer. According to these findings, the most frustrated systems $\theta$-CsCo and $\theta$-CsZn are in a CG state already at the slowest cooling rates feasible in a laboratory experiment, whereas for $\theta$-RbZn and $\theta$-TlZn the first-order CO transition needs to be kinetically avoided by fast cooling in order to prepare the CG state. A more recent combined study of thermal expansion and resistance fluctuation spectroscopy measurements, however, provided evidence for the involvement of structural degrees of freedom \cite{Thomas2022}, at least in the highly frustrated materials $\theta$-CsCo and $\theta$-CsZn. There, a structural glasslike transition at $T_{\mathrm{g}}  = 90 - 100$\,K was found and assigned to the freezing of configurational degrees of freedom of the ET molecules, namely the relative EEG orientation. Moreover, indications for yet another, less strongly pronounced glass transition feature around 120\,K was reported \cite{Thomas2022}, the origin of which has remained unclear.

In the present work, we report the results of thermal expansion measurements on $\theta$-RbZn aiming at investigating potential structural glasslike features in the presence of charge order. Our results, which complement previous findings on the $\theta$-CsCo and $\theta$-CsZn salts \cite{Thomas2022}, both of which lack charge order, reveal clear indications for two glasslike transitions as a common feature in these $\theta$-phase salts, irrespective of the presence or absence of charge order. In addition, we provide model calculations on the glasslike dynamics associated with the flexible EEGs of the ET molecules for various $\theta$-(ET)$_2MM^\prime$(SCN)$_4$ salts. Our calculations support the notion of two different conformational (structural) glass transitions in these salts.

\section{EXPERIMENTAL}
Single crystals of $\theta$-(ET)$_2$RbZn(SCN)$_4$ were grown by electrochemical oxidation \cite{Mori1998}. Typical dimensions of the crystals are about 0.4 $\times$ 0.2 $\times$ 2.5 mm$^3$. Thermal expansion measurements were performed for a single crystal along the $c$ axis that is parallel to the conducting plane from room temperature down to 1.5\,K by using a homemade capacitive dilatometer following the design discussed in Ref.\,\cite{Pott1983}. This technique enables the detection of length changes $\Delta l \geq 10^{-2}$ \AA, where $l$ is the length of the sample. Relative length changes $\Delta l(T)/l$ were measured with $\Delta l(T) = l(T) - l(T_0)$ and $T_0$ the base temperature of the experiment. The obtained $\Delta l(T)/l$ data were differentiated numerically in order to obtain the thermal expansion coefficient $\alpha (T)  = l^{-1} dl/dT$. To this end the $\Delta l(T)/l$ data were divided into equidistant intervals of typically $\Delta T$ = 0.5\,K for temperatures $T\geq$ 12\,K and $\Delta T$ = 0.12\,K for $T\leq$ 12\,K. In each of the intervals the mean slope was determined from a linear regression. The density of data points around 240\,K is reduced due to extrinsic factors giving rise to an enhanced noise level. For hysteresis measurements, special care was taken to ensure cooling or heating at constant sweeping rates $q_\mathrm{c}$ or $q_\mathrm{h}$, respectively, with $|q_\mathrm{c}| = |q_\mathrm{h}|$.

\section{RESULTS}
\subsection{Thermal expansion measurements}
\begin{figure}[htbp]
  \includegraphics[width=8.2cm]{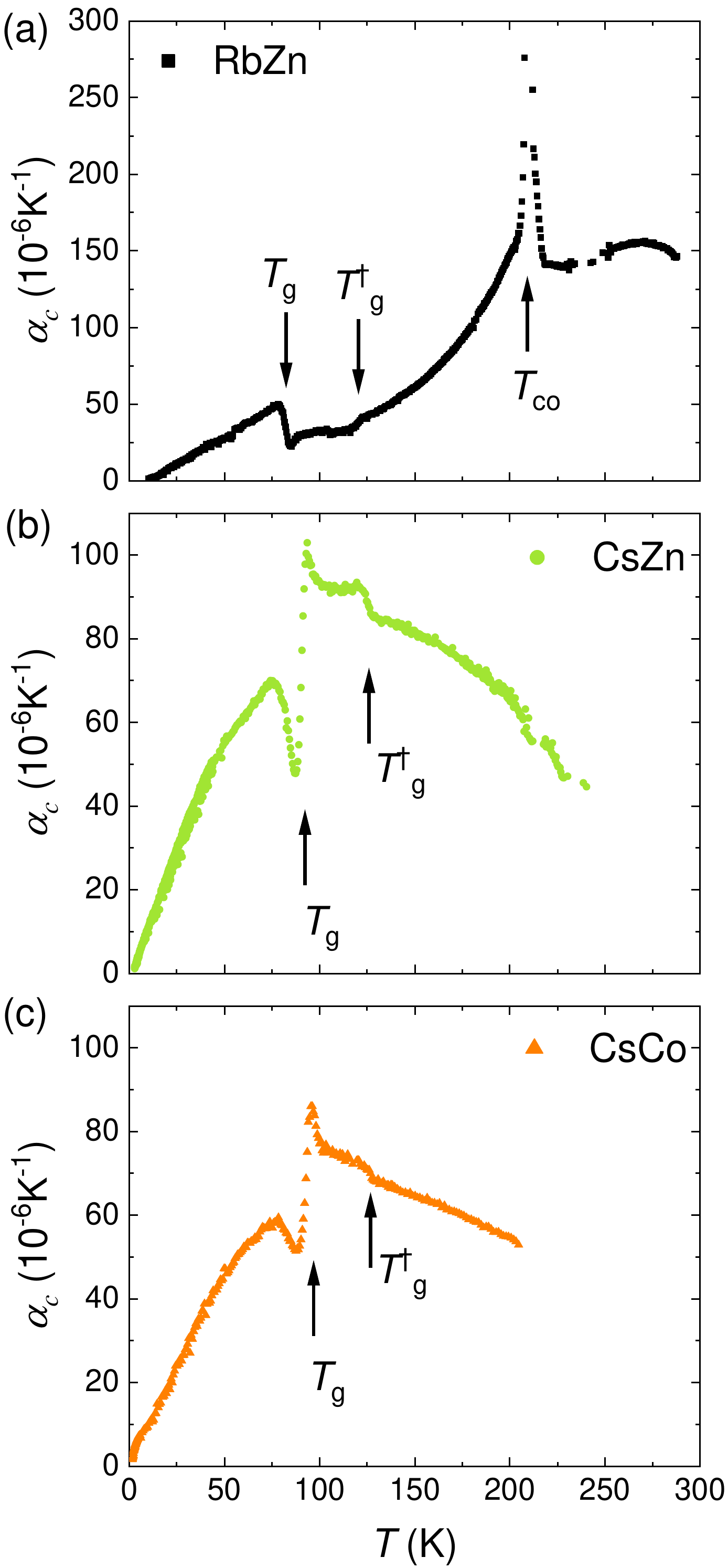}
  \caption{Overview of the thermal expansion coefficient $\alpha_c (T)$ for $\theta$-(ET)$_2MM^\prime$(SCN)$_4$ measured upon heating along the in-plane $c$ axis: (a) $MM^\prime$ = RbZn, (b) $MM^\prime$ = CsZn, and (c) $MM^\prime$ = CsCo. $T_\mathrm{CO}$, $T_\mathrm{g}$, and $T_\mathrm{g}^{\dagger}$ represent the charge-order and glass transition temperatures, respectively. The data in (b) and (c) are taken from Ref.\,\cite{Thomas2022}.}
  \label{Overview}
\end{figure}

\begin{figure}[tbp]
  \includegraphics[width=8.6cm]{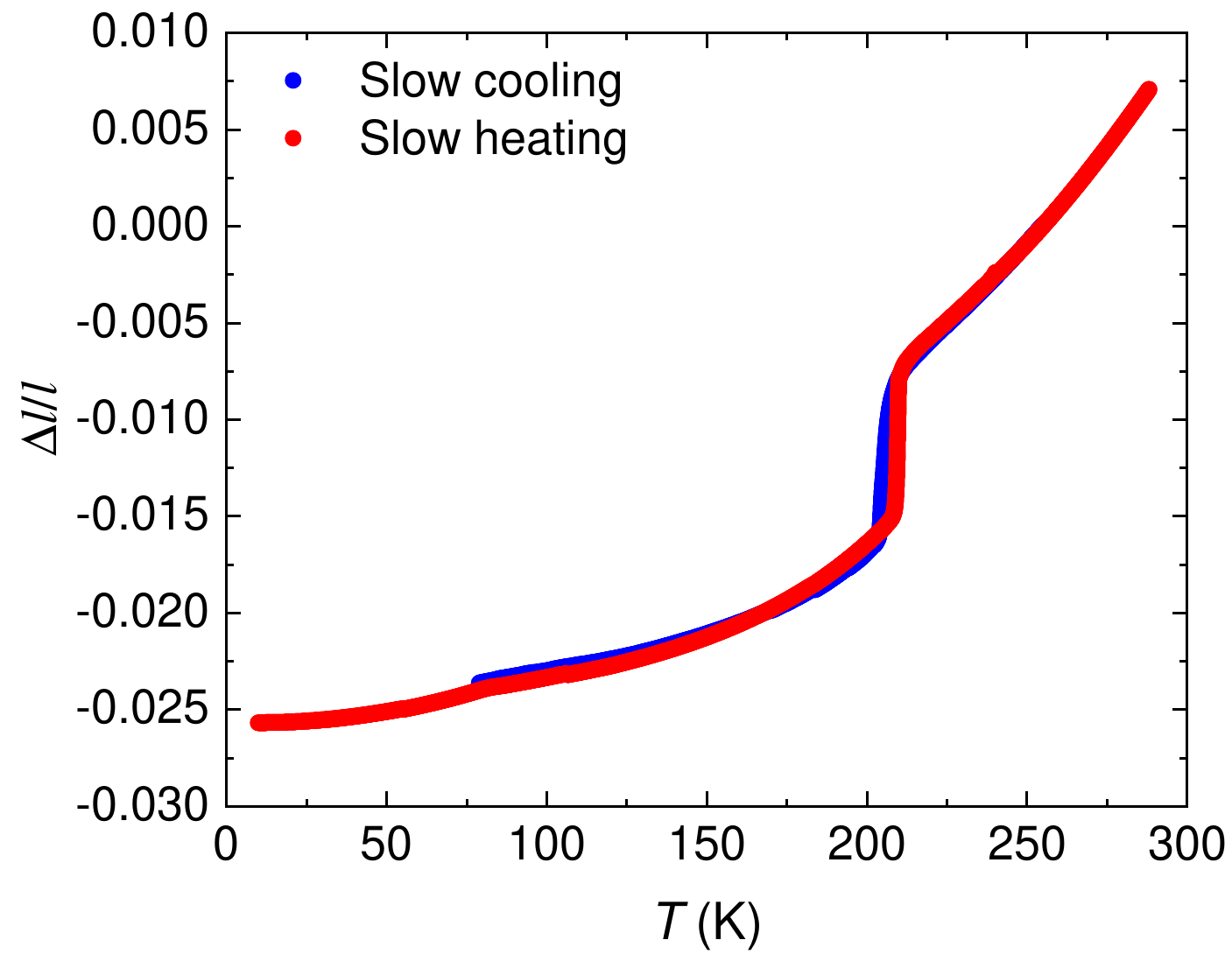}
  \caption{Relative length change for $\theta$-(BEDT-TTF)$_2$RbZn(SCN)$_4$ measured along the $c$ axis upon cooling and heating.}
  \label{Relative_length_change}
\end{figure}
\begin{figure}[bp]
  \includegraphics[width=8.6cm]{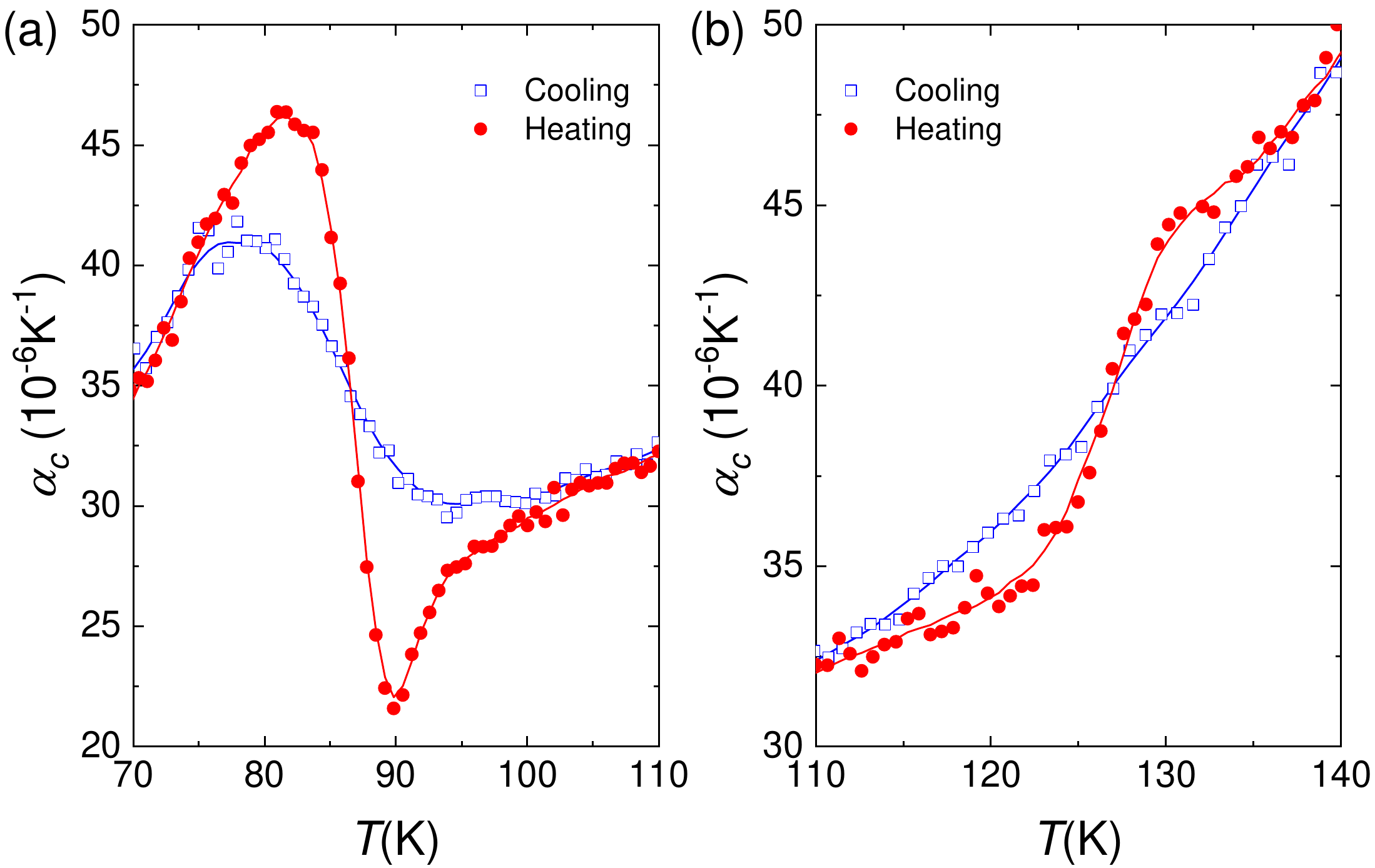}
  \caption{Hysteresis of the thermal expansion coefficient $\alpha_c (T)$ between cooling and heating measurements. Solid lines, representing smoothed curves, are guides to the eye.}\label{Hysteresis}
\end{figure}
\begin{figure}[tbp]
  \includegraphics[width=8.6cm]{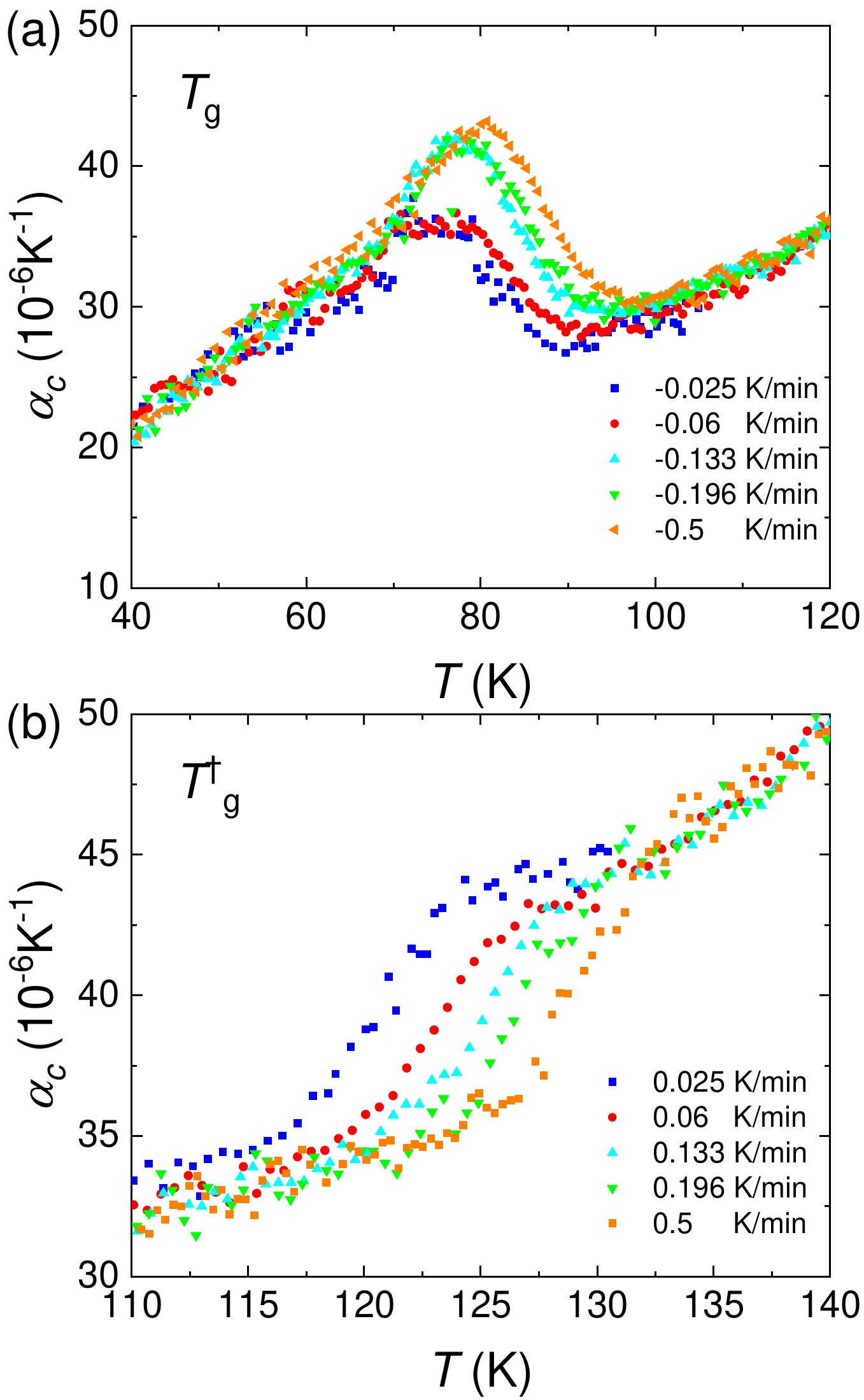}
  \caption{(a) Cooling- and (b) heating-rate dependence of the thermal expansion coefficient $\alpha_c (T)$ for $\theta$-RbZn in the vicinity of the glass transitions at $T_\mathrm{g}$ and $T_\mathrm{g}^{\dagger}$, respectively.}
  \label{cooling_rate_90K}
\end{figure}
Figure \ref{Overview}(a) gives an overview of the thermal expansion coefficient $\alpha (T)$ measured along the in-plane $c$ axis for 1.5\,K $\leq T \leq$ 290\,K. Data were taken upon heating with a rate $q_\mathrm{h} \leq$ 1\,K/min. For comparison, we also show corresponding $\alpha_c$ data for $\theta$-CsZn [Fig.\,\ref{Overview}(b)] and $\theta$-CsCo [Fig.\,\ref{Overview}(c)], reported in Ref.\,\cite{Thomas2022}.

The results for $\theta$-RbZn highlight several remarkable features. (1) $\alpha_c$ is very large around room temperature, reaching values of about 150 $\times$ 10$^{-6}$K$^{-1}$. (2) Upon cooling, a pronounced phase transition anomaly is observed at 215\,K. The corresponding $\Delta l(T)/l$ data, shown in Fig.\,\ref{Relative_length_change}, yield discontinuous length changes at the transition and a small hysteresis upon cooling and warming, clear indications for the first-order character of the transition. We assign this transition to CO accompanied by structural changes as reported in the literature \cite{Mori1998, Watanabe2004}.
(3) Upon further cooling steplike anomalies are observed around 120\,K and around 90\,K. For a closer inspection of these anomalies, we show in Fig.\,\ref{Hysteresis} $\alpha_c$ data taken upon cooling and heating within a narrow temperature interval around $T_{\mathrm{g}}$ and $T_{\mathrm{g}}^\dagger$. While the cooling and heating curves coincide at temperatures sufficiently below and above the respective anomalies, a distinct hysteresis is observed with under- and overshoot behavior in the heating curves characteristic for a structural glasslike transition. Very similar behavior was revealed also for $\theta$-CsZn [Fig.\,\ref{Overview}(b)] and $\theta$-CsCo [Fig.\,\ref{Overview}(c)] in Ref.\,\cite{Thomas2022} where it was assigned to glasslike transitions at $T_{\mathrm{g}} \sim$ 90\,K and $T_{\mathrm{g}}^\dag \sim$ 120\,K, as well as $\kappa$-(ET)$_2$Cu[N(CN)$_2$]$Z$, with $Z$ = Cl and Br where structural glasslike transitions occur at $T_{\mathrm{g}} \sim$ 70\,K and 75\,K, respectively \cite{Muller2002,Muller2004}.

Following the procedure to characterize the glasslike transitions for the $\theta$-CsCo and $\theta$-CsZn salts as well as for other glass-forming materials in Refs.\,\cite{Gugenberger1992,Nagel2000,Muller2002,Thomas2022}, we measured the cooling/heating-rate dependence of the anomalies in $\alpha_c$ for $\theta$-RbZn, for cooling/heating rates 0.025\,K/min $\leq |q_\mathrm{c,h}| \leq$ 0.5\,K/min, see Fig.\,\ref{cooling_rate_90K} (a) and (b). Whereas for the anomaly around 90\,K the evolution of $T_{\mathrm{g}}$ can be followed by analyzing the cooling curves, heating curves are used for the anomaly around 120\,K for ease of analysis by tracking the steplike anomaly seen there. $T_{\mathrm{g}}$ and $T_{\mathrm{g}}^{\dagger}$ are defined as the midpoint of the steplike feature in the cooling or heating curves, respectively.

\begin{figure}[tbp]
  \includegraphics[width=8.6cm]{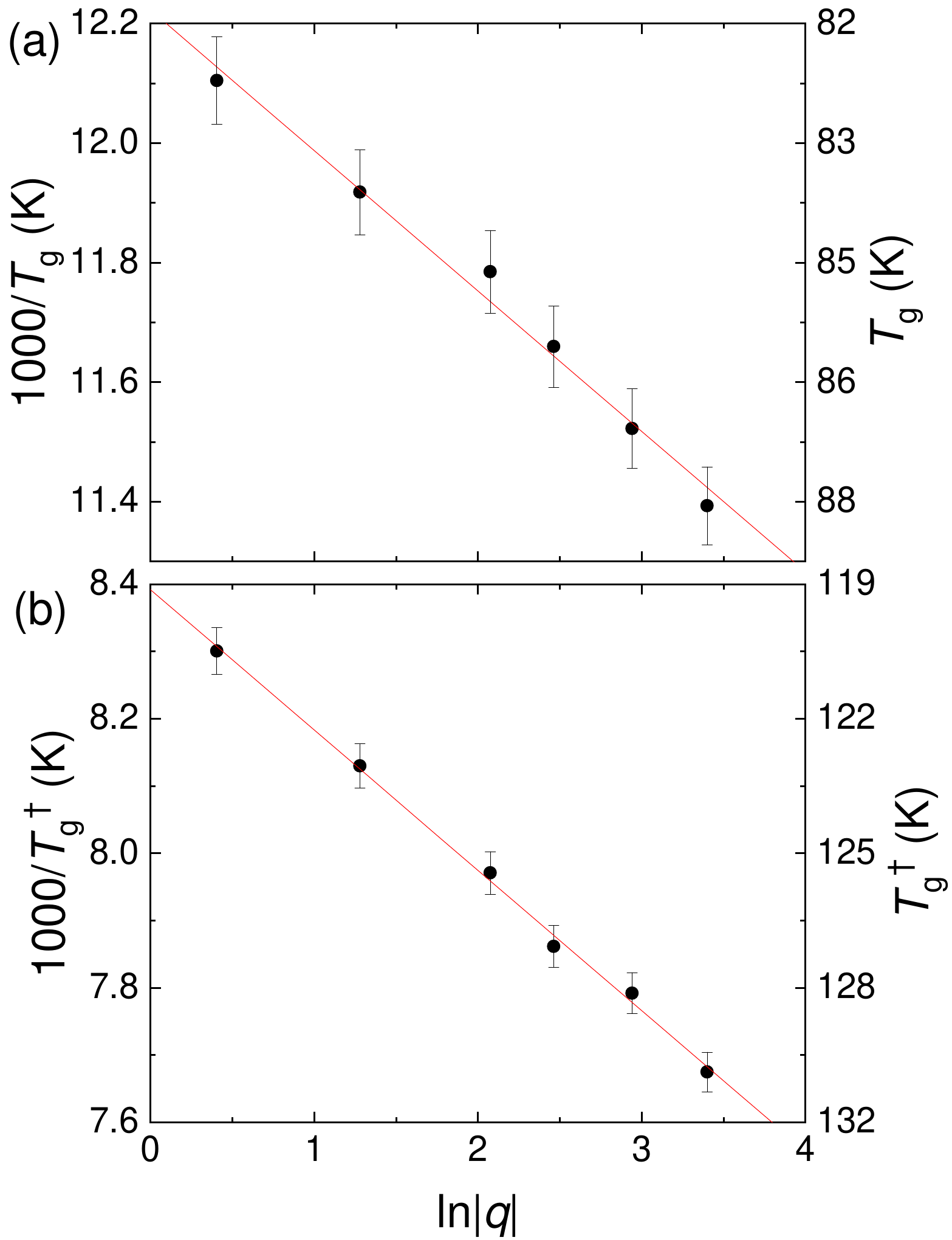}
  \caption{Arrhenius plot of (a) $T_{\mathrm{g}}^{-1}$ vs. $|q_\mathrm{c}|$ and (b) $T_{\mathrm{g}}^{\dagger -1}$ vs. $|q_\mathrm{h}|$ for $\theta$-RbZn.}
  \label{Activation_energy_90K}
\end{figure}

With increasing the cooling/heating rate, the glass transition temperatures shift to higher temperatures as expected for a glass-forming system in which the relaxation time $\tau$ increases with lowering the temperature. To analyze this shift quantitatively, we show in Fig.\,\ref{Activation_energy_90K} (a) and (b) an Arrhenius plot of $T_{\mathrm{g}}^{-1}$ vs. the cooling rate $|q_\mathrm{c}|$ and $T_\mathrm{g}^{\dagger}$ vs. the heating rate $q_\mathrm{h}$, respectively (see Refs.\,\cite{Muller2002, Thomas2022} for details). The linear behaviors revealed in this representation indicate thermally activated relaxation times for both cases $\tau \propto \exp [E_\mathrm{a}/(k_\mathrm{B}T)]$, where $E_\mathrm{a}$ represents an activation energy. Within a simple two-level model one finds $\ln |q_\mathrm{c,h}| = - E_\mathrm{a}/(k_\mathrm{B}T) + const.$ \cite{Nagel2000,Muller2002}. Linear fits to the data in Fig.\,\ref{Activation_energy_90K} yield $E_\mathrm{a}/k_\mathrm{B}$ = (4250 $\pm$ 250)\,K for $T_{\mathrm{g}}^{-1}$ and $E_\mathrm{a}^{\dagger}/k_\mathrm{B}$ = (4800 $\pm$ 130)\,K for $T_{\mathrm{g}}^{\dagger -1}$.

\subsection{CALCULATIONS}
\begin{table*}[tb]
\caption{Computed energetics for flipping of ethylene endgroups (EEGs), according to the definitions in Fig.~\ref{Transition_diagram}.  Computed values were averaged over different conformations of neighboring EEGs (see text). Each quantity $Q$ is given as $Q \pm SD$, where $SD$ is one standard deviation of the computed energies for different neighboring EEG conformations. }
\begin{tabular}{|l|llllllll}
\hline
      & \multicolumn{3}{c|}{Facing Rb/Cs}   & \multicolumn{3}{c|}{Facing Zn/Co}       \\ \hline
$MM'$   & \multicolumn{1}{c|}{$E_\mathrm{a}/k_\mathrm{B}$ (K)} & \multicolumn{1}{c|}{$\Delta E/k_\mathrm{B}$ (K)} & \multicolumn{1}{c|}{$E_\mathrm{a} /\Delta E$}   & \multicolumn{1}{c|}{$E_\mathrm{a}/k_\mathrm{B}$ (K)} & \multicolumn{1}{c|}{$\Delta E/k_\mathrm{B}$ (K)} & \multicolumn{1}{c|}{$E_\mathrm{a} /\Delta E$} \\ \hline

RbCo  & \multicolumn{1}{l|}{4600 $\pm$ 590}      & \multicolumn{1}{l|}{570 $\pm$ 280}            & \multicolumn{1}{c|}{10}  & \multicolumn{1}{l|}{5660 $\pm$ 650}      & \multicolumn{1}{l|}{200 $\pm$ 300}            & \multicolumn{1}{c|}{23}     \\ \hline
RbZn  & \multicolumn{1}{l|}{4340 $\pm$ 600}      & \multicolumn{1}{l|}{580 $\pm$ 300}              & \multicolumn{1}{c|}{11}  & \multicolumn{1}{l|}{5820 $\pm$ 630}      & \multicolumn{1}{l|}{150 $\pm$ 300}            & \multicolumn{1}{c|}{29}        \\ \hline
CsCo  & \multicolumn{1}{l|}{4270 $\pm$ 360}      & \multicolumn{1}{l|}{140 $\pm$ 150}             & \multicolumn{1}{c|}{33}  & \multicolumn{1}{l|}{5510 $\pm$ 340}      & \multicolumn{1}{l|}{120 $\pm$ 130}            & \multicolumn{1}{c|}{62}      \\ \hline
CsZn  & \multicolumn{1}{l|}{3800 $\pm$ 350}      & \multicolumn{1}{l|}{170 $\pm$ 160}              & \multicolumn{1}{c|}{29}   & \multicolumn{1}{l|}{5050 $\pm$ 290}      & \multicolumn{1}{c|}{2 $\pm$ 130}    & \multicolumn{1}{c|}{85}      \\ \hline
\end{tabular}
\label{Activation_energies_in-stack}
\end{table*}

\begin{figure}
  \includegraphics[width=6cm]{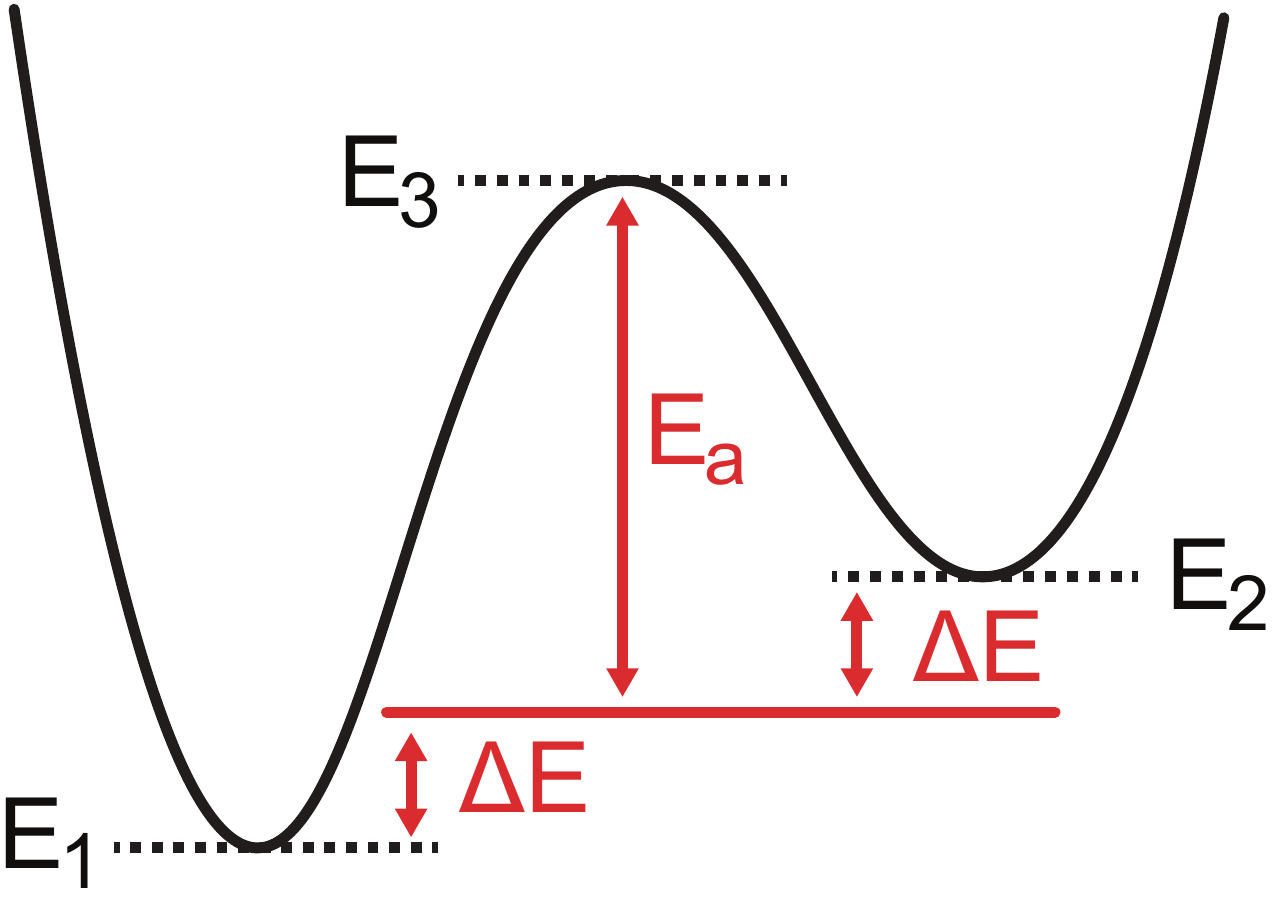}
  \caption{Transition diagram. Energies $E_1$ and $E_2$ indicate the two low energy states and $E_3$ is the high energy transition state between them. $E_2 - E_1 = 2\Delta E$, and $E_\mathrm{a} = E_3-E_1+\Delta E$ denotes the activation energy. }\label{Transition_diagram}
\end{figure}

In order to elucidate the nature of the multiple observed glass transitions, we performed {\it ab-initio} calculations of the energy landscape for EEG dynamics. In particular, we followed an approach inspired by the computational study in Ref.~\cite{Muller2015}, in order to facilitate comparison with other structural glass-forming $\kappa$-phase ET salts. In reference to the simplified diagram in Fig.~\ref{Transition_diagram}, it was previously observed that a computed ratio $E_\mathrm{a} / \Delta E \gtrsim 5$ was an empirical requirement for glassy freezing of the EEGs. Large ratios of $E_\mathrm{a} / \Delta E$ promote metastability of different conformations, as it implies the energy barrier for relaxation between conformations is large compared to their relative energy differences. We therefore sought to estimate the relevant energies for different EEG conformations in the present materials.

All Density Functional Theory (DFT) calculations were carried out with ORCA 5.0 \cite{neese2022software} at the B3LYP/def2-SV(P) level. We obtained starting geometries from room temperature crystal structures for $\theta$-(BEDT-TTF)$_2MM^\prime$(SCN)$_4$ ($M$ = Rb, Cs, $M^\prime$ = Zn, Co) materials from Refs.~\cite{Mori1998a,Mori1995}.
Interactions between the EEGs and nearby anions were modeled with a combination of a Lennard-Jones and electrostatic point-charge potential following the OPLS scheme \cite{damm1997opls}. For C, H, S, and N, we employed the standard non-bonded OPLS Lennard-Jones parameters. For Rb, Cs, Zn, and Co, we adapted parameters from Ref.\,\cite{li2013rational,li2015systematic}. Atomic charges for the anion layers were estimated from the Mulliken charges computed for $M^\prime$(SCN)$_4$ clusters. The atomic charge of the alkali metals Rb and Cs was taken to be +1$e$. In all calculations, the geometry of the central TTF moeities was constrained to be that of the reported crystal structures, and only the coordinates of selected -S$_2$C$_2$H$_4$ EEG units were relaxed. To find saddle-point transition states between two stable geometries, we used ORCA's hybrid Hessian mode-following algorithm \cite{neese2012orca}.

\begin{figure*}[t]
  \includegraphics[width=17cm]{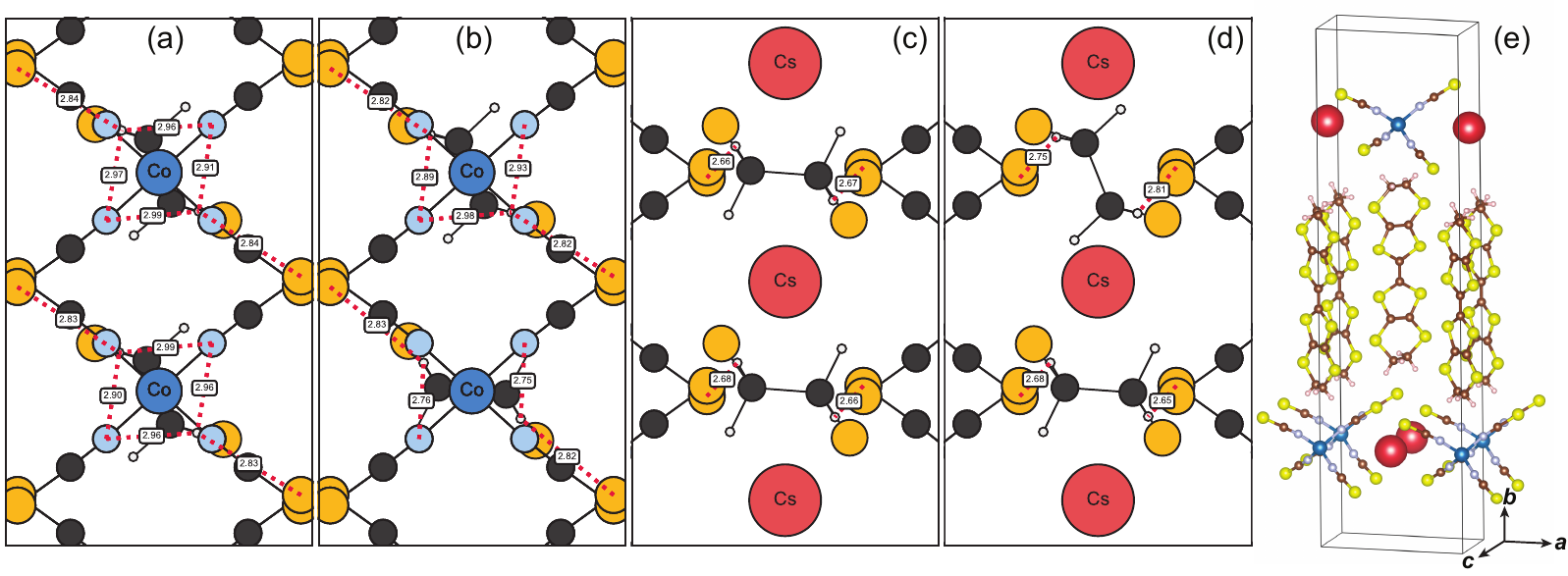}
  \caption{The computed orientation of the EEGs before and after flipping for $\theta$-CsCo viewed along the $b$-axis. (a,c) The lowest energy conformations for Co-facing EEGs and Cs-facing EEGs, respectively. (b,d) Conformations after a flip of a single EEG.
  Hydrogen bonds less than three angstroms in length are drawn in red dotted lines and labelled. (e) View of a single ET layer and neighboring anions for perspective.}
\label{Transition_mechanisms}
\end{figure*}

In contrast with previously studied $\kappa$-phase salts, the $\pi$-stacking of the $\theta$-phase allows for significant cooperativity between multiple EEGs in the same stack. In order to account for this, we employed a two-step procedure to explore the full energy landscape of EEG conformations. For each material, we first computed relaxed geometries for two adjacent ET molecules in a $\pi$-stack in the presence of the anion layer force field including atoms in the anion layer within 16 \AA. There are, in total, $2^4 = 16$ possible EEG conformations per molecular pair (although some are symmetry-related). We refer to the different EEGs according to the atoms in the anion layer they face: towards the Rb/Cs or Zn/Co. We find in all materials that there is an average energetic preference for every ET to be in a staggered conformation with Rb/Cs-facing EEGs oriented horizontal with the $c$-axis, and Zn/Co-facing EEGs oriented at an angle in the $ac$-plane (see Fig.~\ref{Transition_mechanisms}). It may also be noted that the energies for flipping an EEG facing the Rb/Cs do not depend strongly on the conformations of the EEGs on the opposite sides of the molecule facing the Zn/Co (and vice versa). As a consequence the EEGs on opposite sides of the molecule can be treated as independent.

In order to explore the cooperativity between adjacent EEGs on the same side of the ETs, we then computed energies $\Delta E$ and $E_\mathrm{a}$ for flipping a given EEG as a function of it's neighboring EEG conformations. To accomplish this, we took the optimized geometries of the two-ET calculations as starting geometries for calculations including four ETs in the same $\pi$-stack. In each case, we constrain the conformation of ETs on the top and bottom of the stack, and reoptimize the EEGs on the central molecules. For four ETs there are, in total $2^8 = 256$ conformations. In order to reduce the number of computations, we focused on the EEGs on each side of the molecules separately; we computed energies for all 16 conformations and corresponding transition states of the four EEGs facing the Rb/Cs while keeping the Zn/Co-facing EEGs in their lowest energy conformation. This procedure was then repeated for the EEGs facing the Zn/Co, while keeping the Rb/Cs-facing EEGs in their lowest energy conformation. The obtained parameters $\Delta E$ and $E_\mathrm{a}$, averaged over the different conformations, are shown in Table \ref{Activation_energies_in-stack}, along with their standard deviations. The latter quantities reflect the variation of $\Delta E$ and $E_\mathrm{a}$ for flipping a particular EEG as a result of the conformations of their neighbors.

The observations from these calculations are threefold. First we find, for all materials, that the average $\Delta E$ and $E_\mathrm{a}$ values satisfy $E_\mathrm{a} / \Delta E \gtrsim 5$ for both types of symmetry-inequivalent EEGs. In all conformations, the EEGs are able to form close H$\cdot\cdot\cdot$SCN and H$\cdot\cdot\cdot$NCS contacts with the anion layer, so that no conformation is strongly stabilized by electrostatic and van der Waals interactions with the anions.
This is empirically consistent with both the Rb/Cs-facing and Zn/Co-facing EEGs exhibiting glassiness. Furthermore, the two types of EEGs are essentially energetically decoupled, lending significant support to the interpretation of the thermal expansion anomalies as two separate and distinct structural glass transitions. In practice, longer range strain effects (not considered here) may result in a coupling of the two types of EEGs, but such effects are evidently not strong enough to merge the two glass transitions.

Second, we find that the computed activation energies for EEG conformational changes are compatible with the experimental values obtained above and in Ref.~\onlinecite{Thomas2022} from the cooling-rate dependence of $T_{\mathrm{g}}$ and $T_\mathrm{g}^{\dagger}$. The Rb/Cs-facing EEGs are computed to have activation energies in the range $E_a/k_\mathrm{B} \sim$ 3800 - 4600\,K, while the Zn/Co-facing EEGs are computed to have $E_\mathrm{a}/k_\mathrm{B} \sim$ 5000 - 5900\,K. These may be compared with the experimental values $E_\mathrm{a}/k_\mathrm{B} = (4250 \pm 250)$\,K and $E_\mathrm{a}^\dagger/k_\mathrm{B} = (4800 \pm 130)$\,K (for $\theta$-RbZn), $E_\mathrm{a}/k_\mathrm{B} = (3900 \pm 170)$\,K and $E_\mathrm{a}^\dagger/k_\mathrm{B} = (4900 \pm 350)$\,K (for $\theta$-CsZn, $\theta$-CsCo)\cite{Thomas2022}. We thus identify $T_{\mathrm{g}}$ with the freezing of the Rb/Cs-facing EEGs, and $T_\mathrm{g}^{\dagger}$ with the freezing of the Zn/Co-facing EEGs. Again, the computational studies support the identification of the two thermal expansion anomalies as distinct structural glass transitions.

\begin{figure*}[tbp]
  \includegraphics[width=15cm]{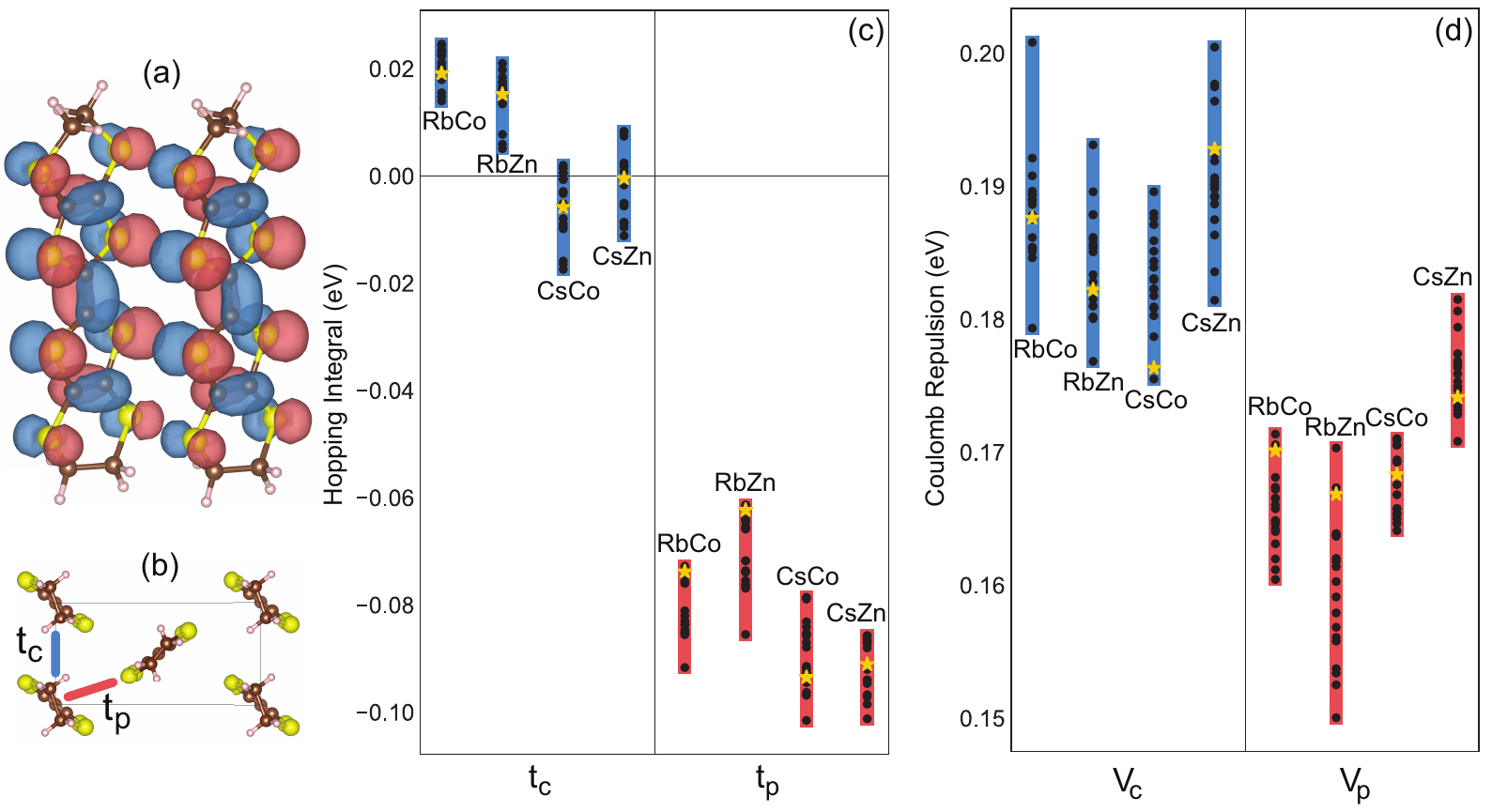}
  \caption{(a) Highest occupied molecular orbitals for two adjacent molecules in a $\pi$-stack. (b) Unit cell viewed along the $b$-axis showing definition of $t_p$ and $tc$ hopping integrals. (c,d) Range of computed hopping integrals $t_{ij}$ and Coulomb repulsion $V_{ij}$, with reference to the Hamiltonians $\sum_{ij,\sigma}t_{ij} c_{i,\sigma}^\dagger c_{j,\sigma}$ and $\sum_{ij} V_{ij} n_i n_j$, respectively. Each dot in a vertical stack corresponds to a different conformation (16 in total for each pair of ETs). The vertical lines indicate the range of the computed parameters. The yellow star indicates the value for the lowest energy conformation.}
  \label{Hopping_integrals}
\end{figure*}
Third, it should be emphasized that the spreads of $\Delta E$ values for flipping a single EEG as a function of the conformation of adjacent EEGs were found to be similar in magnitude to the $\Delta E$ values themselves. This implies (a) that the ultimate energetic balance between different conformations is sensitive, and may be influenced by additional factors such as charge order, and (b) there is significant steric interaction between EEGs on adjacent ET molecules in the same $\pi$-stack. Such interactions lead to cooperative dynamics in glass formers, which may be evidenced by growth of the relaxation times on cooling that proceeds faster than Arrhenius behavior (see \cite{bohmer1993nonexponential,angell1995formation,lunkenheimer2000glassy,bauer2013cooperativity} and references therein). On the basis of significant steric coupling, we anticipate that $\theta$-phase salts fall into the category of 'fragile' glass formers, similar to their $\kappa$-phase counterparts \cite{Muller2015}. We note that a signature of fragile glasses is a deviation from an Arrhenius behavior, mostly seen as a curvature in the relaxation time upon approaching the glass transition temperature in a plot $\log{\tau}$ vs.\ $T_g^{-1}$. This is apparently not seen for the cooling-/heating-rate dependent anomalies in thermal expansion capturing the static glass transition shown in Fig.\,\ref{Activation_energy_90K}, similar as for the glassy anomalies in the $\kappa$-phase salts \cite{Muller2002}, because the 'frequency range' or the window of relevant relaxation times determined by $|q_{c,h}|$ is too narrow. In complementary studies of the \textit{dynamic} glass transition by fluctuation (noise) spectroscopy of the same $\theta$- and $\kappa$-phase materials, where a much wider frequency range is covered, a clear curvature described by a Vogel-Fulcher-Tammann behavior, a signature of cooperativity and a 'fragile' glass-forming system, is observed \cite{Thomas2022,Muller2015}.

Finally, the identification of structural glassiness in \mbox{$\theta$-$MM^\prime$} compounds raises the question of the potential impact of the structure on the charge dynamics. In order to address this question, we estimated the nearest neighbor hopping integrals $t_{ij}$ and Coulomb repulsions $V_{ij}$ as a function of EEG conformation. There are two types of symmetry distinct nearest neighbors: intrastack neighbors (labelled $t_c$, $V_c$), and interstack neighbors (labelled $t_p$, $V_p$). We employed the optimized geometries of the central molecules from the 4-ET calculations. The pairwise hoppings between ET molecules were computed from the DFT Fock matrix using ORCA according to the method outlined in \cite{ganter2022database}. To estimate $V_{ij}$, we employed the approach in \cite{Mori2000V}:
\begin{align}
V_{ij} \approx e^2\sum_{a,b}\frac{|\phi_{i,a}(\mathbf{r}_a)|^2|\phi_{j,b}(\mathbf{r}_b)|^2}{|\mathbf{r}_a-\mathbf{r}_b|}
\end{align}
where $\phi_{i,a}$ is the computed molecular orbital coefficient of the atom $a$, located at position $\mathbf{r}_a$ for the HOMO of ET molecule $i$. This approximation effectively treats each electron as a distribution of point charges centered at the atoms in each ET. We note that this approach \emph{underestimates} the structural dependence of the electronic parameters, because the geometries of the TTF cores of each molecule are constrained during the geometry optimizations, so effects of the EEG orientations alone are considered.  Results are shown in Fig. \ref{Hopping_integrals}. Consistent with experiments, we find $\theta$-CsZn and $\theta$-CsCo to have a higher average degree of charge frustration than their Rb counterparts, as indicated by the average ratio $V_c/V_p$ closer to 1. However, both the hopping integrals and Coulomb repulsion terms show a substantial degree of dependence on the conformation, varying by 10 - 20\%. To the extent that such variations may alter the local degree of  charge frustration, this finding suggests that the electronic degrees of freedom may be relatively strongly coupled to the EEG conformations. This raises the possibility that the glassy freezing of EEGs, leading to quenched disorder in the electronic parameters, could modify or enhance glassy charge dynamics below the structural glass transition.

\section{DISCUSSION}
\subsection{Charge-order transition}
The discontinuous length changes and hysteretic behavior revealed for $\theta$-RbZn along the $c$ axis around 215\,K (Fig.\,\ref{Relative_length_change}), giving rise to a pronounced anomaly in $\alpha_c(T)$ [Fig.\,\ref{Overview}(a)], indicate a first-order phase transition. In this temperature range, $\theta$-RbZn is known to undergo a transition from a high-temperature metallic state to a low-temperature charge-ordered insulating state accompanied by a lattice modulation \cite{Mori1998,Watanabe2004,Sato2020}. Accordingly, we assign this feature to the signature of the combined CO/structural transition. Generally, due to the ionic character of the organic conductors, ordering in the charge sector is expected to be accompanied by lattice effects, see, e.g., the anomalies revealed in the coefficient of thermal expansion at the CO transition for various (TMTTF)$_2X$ salts \cite{DeSouza2008} and $\kappa$-(ET)$_2$Hg(SCN)$_2$Cl \cite{Gati2018b}. Apart from the phase transition anomaly, Fig.\,\ref{Overview} highlights a remarkable difference both in the size and temperature dependence of $\alpha_c(T)$ at $T >$ 100\,K for $\theta$-RbZn as opposed to $\theta$-CsZn and $\theta$-CsCo. Whereas $\alpha_c(T)$ for $\theta$-RbZn is extraordinarily large, reaching values around 150 $\times$ 10$^{-6}$K$^{-1}$, and shows the usual reduction upon cooling, a distinctly different behavior is seen for $\theta$-CsZn and $\theta$-CsCo, where $\alpha_c(T)$ around 200\,K is distinctly smaller and grows upon cooling, reaching a maximum around 100\,K. Interestingly enough, these marked differences in the high-temperature thermal contraction for the $M$ = Rb salt vs.\ the $M$ = Cs salts appear to have no effect on the occurrence (and position) of the glasslike transitions at $T_{\mathrm{g}}$ and $T_{\mathrm{g}}^\dag$.

A remarkable feature reported for the $\theta$-RbZn salt is that the CO and accompanied structural transition can be kinetically avoided when the crystal is cooled faster than a critical cooling rate $q_\mathrm{c}^{\rm crit}$. In Refs.\,\cite{Nogami2010,Kagawa2013,Sato2017} $|q_\mathrm{c}^{\rm crit}|$ was found to be around 5\,K/min. As shown in Fig.\,\ref{Relative_length_change2} in the Appendix, our attempt to kinetically avoid the transition by fast cooling with a rate of about 10\,K/min, the maximum cooling rate accessible by our dilatometer around 200\,K, failed. The data obtained indicate a shift of the transition $T_{\mathrm{CO}}$ to a lower temperature of about 170\,K  accompanied by a significant broadening. This may indicate a sample-to-sample variation in the critical cooling rate. Our finding in thermal expansion measurements is very similar to the observation in Ref.\,\cite{Thomas2022b}, where also for cooling rates of $q_\mathrm{c}^{\rm crit} = 10$\,K/min the sharp first-order transition was not completely suppressed and a broadened, step-like anomaly in the resistivity, shifted to lower temperatures, was still observed. This indicates that the critical cooling rate of $q_\mathrm{c}^{\rm crit} = 5$\,K/min reported in the literature \cite{Kagawa2013} is too slow, which we attribute to the higher quality of the present samples as discussed in Ref.\,\cite{Thomas2022b}.

\subsection{Glasslike transition at 80--90 K}
Cooling a material from a liquid state into a glassy state is a non-equilibrium process by which the components involved continue to change over time with a characteristic relaxation time $\tau$. Once $\tau$ becomes greater than the experimental observation time $\Delta t$ with increasing $|q_\mathrm{c}|$, the relaxation cannot be completed before the temperature further decreases. As a result, depending on $|q_\mathrm{c}|$, the involved structural components fall out of equilibrium and become frozen in a glassy state below $T_{\mathrm{g}}$($q_\mathrm{c}$). In a thermal expansion measurement this cooling process is accompanied by a broadened and rounded step-like anomaly $\Delta \alpha_i$. Since $\Delta \alpha_i$ is related to the uniaxial-pressure dependence of the entropy by $\Delta \alpha_i = -\frac{1}{V_{\rm mol}} \frac{\partial S_{\rm EEG}}{\partial p_i}|_T$ for $i = a,b,c$, where $S_{\rm EEG}$ is the molar entropy associated with the EEG disorder \cite{Nagel2000} and $V_\mathrm{mol}$ the molar volume, $\Delta \alpha_i$ can be positive or negative. As an example, we mention the highly anisotropic glasslike transition in $\kappa$-(ET)$_2$Cu[N(CN)$_2$]Br \cite{Muller2004}.
In contrast, upon heating, more sharp and well-pronounced step-like features are observed which are accompanied by characteristic over- and undershoot behavior yielding positive and negative jumps, depending on the direction \cite{Gugenberger1992,Nagel2000,Muller2002,Muller2004}, i.e., there is a pronounced hysteresis the width of which is characteristic to the energetics of the transition. 
Accordingly, the steplike changes in $\alpha_i$ upon heating through $T_{\mathrm{g}}$, along with the occurrence of hysteresis around $T_{\mathrm{g}}$ and the heating/cooling-rate dependence of $T_{\mathrm{g}}$, can be used as defining criteria to identify glasslike transitions via thermal expansion measurements \cite{Gugenberger1992, Nagel2000, Muller2002, Gati2018}.

The data presented in Fig.\,\ref{cooling_rate_90K}(a) thus provide clear evidence for a glasslike transition at $T_{\mathrm{g}}$ in $\theta$-RbZn. The results are qualitatively similar to those found in $\theta$-CsZn and $\theta$-CsCo at 90-100\,K. For the latter two salts, the glasslike anomaly was assigned to the freezing of structural conformations of the ethylene endgroups in the ET molecules \cite{Thomas2022} in analogy to the observations made for the various $\kappa$-type (ET)$_2X$ salts \cite{Miyagawa1995, Tanatar1999, Akutsu2000, Muller2002}. To get deeper insight into the nature of the anomaly for $\theta$-RbZn, we determined the corresponding activation energy. As shown above, from the linear fit to the data in Fig.\,\ref{Activation_energy_90K}(a), we obtain $E_\mathrm{a}/k_\mathrm{B}$ = (4250 $\pm$ 250)\,K. This activation energy is of similar size as $E_\mathrm{a}/k_\mathrm{B}$ = (3750 $\pm$ 150)\,K and (4080 $\pm$ 240)\,K obtained from thermal expansion measurements on $\theta$-CsZn and $\theta$-CsCo, respectively, indicating that the ethylene endgroups are the relevant entities for the relaxation process also in $\theta$-RbZn. This conclusion is also corroborated by the results of our model calculations, see subsection \ref{model calculations} below. An interesting observation made in the present study relates to the sign of $\Delta \alpha_i = \alpha_i(T \rightarrow T_\mathrm{g}^{+}) - \alpha_i(T \rightarrow T_\mathrm{g}^{-})$ at $T_{\mathrm{g}}$ (and $T_{\mathrm{g}}^\dag$, see below). Whereas $\Delta \alpha_c <$ 0 for $\theta$-RbZn, we found $\Delta \alpha_c >$ 0 for $\theta$-CsCo and $\theta$-CsZn. This implies that uniaxial pressure along the $c$ axis decreases the degree of EEG order for $\theta$-RbZn, whereas it increases the degree of order for $\theta$-CsCo and $\theta$-CsZn. As a possible explanation we suggest that the behavior may depend on whether or not the system is in the CO state, implying a structural change from orthorhombic $I$222 to $P$2$_1$2$_1$2$_1$, when the transition occurs at $T_\mathrm{CO}$ \cite{Watanabe2004}.

\subsection{Glasslike transition at 120 -- 130 K}
In $\theta$-CsZn and $\theta$-CsCo, resistance fluctuation (noise) spectroscopy shows that the energy and spectral weight distribution of an enhanced noise level at 175\,K ($E_\mathrm{a}^\dag/k_\mathrm{B}$ = 4990\,K) matches very well the activation energy of the cooling-rate-dependent anomaly at $T_{\mathrm{g}}^\dag (q_\mathrm{c}) \sim$ 120-130\,K in the thermal expansion coefficient, which implies that the slowing down of the charge dynamics observed at 175\,K at a frequency of $f = 1$\,Hz is related to the glasslike transition at about $T_{\mathrm{g}}^\dag$ seen in 'static' thermal expansion measurements for different cooling/warming rates \cite{Thomas2022}. It has been pointed out that this anomaly at around 120-130\,K coincides with a minimum in the in-plane resistivity \cite{Thomas2022}, indicating a crossover from metallic to semiconducting behavior of the 2D-confined electron fluid \cite{Sato2020}. Moreover at about the same temperature the development of a superlattice structure with wave vector $q_1 \sim$ (2/3, $k$, 1/3) was observed by x-ray diffuse scattering \cite{Nogami1999,Watanabe1999,Sato2014} and has been interpreted as the growth and subsequent freezing of charge clusters upon cooling.

An important finding of the present work is that in the $\theta$-RbZn salt there is the same type of anomaly at $T_{\mathrm{g}}^\dag$ around 120\,K as for the $\theta$-CsZn and $\theta$-CsCo counterparts, where the activation energy obtained from thermal expansion measurements of $E_\mathrm{a}^{\dagger}/k_\mathrm{B}$ = (4800 $\pm$ 130)\,K for $\theta$-RbZn, is of similar size as  (5440 $\pm$ 480)\,K for $\theta$-CsZn, and (4920 $\pm$ 400)\,K for $\theta$-CsCo, also revealed by thermal expansion measurements. Thus, the occurrence of this anomaly, regardless of the presence ($\theta$-RbZn) or absence ($\theta$-CsZn and $\theta$-CsCo) of charge order, rules out that the structural glassiness is caused by the charge-glass formation.
Rather, the glassy EEG freezing seems to be a common feature for the $\theta$-(ET)$_2$$MM’$(SCN)$_4$ family. Finally, we like to point out that the sign of this anomaly at $T_{\mathrm{g}}^\dag$ shares the same systematics as revealed for the anomaly at $T_{\mathrm{g}}$, i.e., opposite sign to the corresponding features seen in $\theta$-CsCo and $\theta$-CsZn: We find $\Delta \alpha_c > 0$ for $\theta$-RbZn, as opposed to $\Delta \alpha_c < 0$ for $\theta$-CsCo and $\theta$-CsZn.

\subsection{Comparison with model calculations}\label{model calculations}
On discussing possible origins for the second glasslike transition at $T_{\mathrm{g}}^\dag$ it is worth mentioning that two glassy transitions were also observed for $\kappa$-(ET)$_2$Cu(NCS)$_2$ \cite{Muller2002,Kuwata2011}. For this salt there are two crystallographically independent dimers of ET molecules that can be distinguished by NMR \cite{Saito2015}, such that the ethylene motion may freeze independently. In the present $\theta$-(ET)$_2MM’$(SCN)$_4$ salts there is only one crystallographically independent ET molecule outside of the charge-ordered phase. However, the EEGs on different ends of each molecule are inequivalent, allowing for two distinct structural glass transitions. Consistently, we found that all $E_\mathrm{a}/\Delta E$ values for the EEGs facing both Zn/Co as well as Rb/Cs atoms are greater than five (see Table \ref{Activation_energies_in-stack}), which was previously identified as the empirical threshold value above which glasslike behavior occurs \cite{Muller2015}. The activation energies governing these dynamics were estimated to fall in the experimentally observed range, with $E_\mathrm{a}$ for the Zn/Co-facing EEGs being consistently larger than for the Rb/Cs-facing EEGs. This finding supports the notion that two different conformational glass transitions are possible in this family of materials. We assign the glasslike anomaly at 120-130\,K to the freezing of the EEGs facing the Zn/Co and the glasslike anomaly at 90-100\,K to the freezing of the EEGs facing the Rb/Cs.

We note that a mechanism related to the ordering of the EEGs was proposed in Ref.\,\cite{Alemany2015} to account for the charge ordering metal-insulator transition in $\theta$-RbZn. It was argued that the transition is an order-disorder structural transition of the EEGs doubling the periodicity along $c$ --- the stack direction --- and driving the system into an electronically pseudo-1D situation along $a$ --- the interstack direction. As the freezing occurs well below the metal-insulator transition, our results point against such a structural mechanism for the charge order transition.

Finally, it is important to consider possible implications for $\theta$-CsZn and $\theta$-CsCo, where long-range charge order is avoided. Our calculations of the hopping integrals and Coulomb repulsion terms show a potential for quenched disorder in the electronic parameters on the scale of 10-20\% due to the glassy freezing of the EEGs. In conjunction with the frustration of charge order due to the triangular arrangement of the ET molecules, such disorder may be sufficient to induce glassy charge dynamics in these compounds by locally favoring different charge ordering patterns. Consistently, the onset of diffuse superlattice scattering in $\theta$-CsZn, indicating the formation of local charge clusters, occurs only below $\sim$ 100 K \cite{Sato2014}. Given that this temperature corresponds with the now-identified structural $T_{\mathrm{g}}$ and $T_{\mathrm{g}}^\dagger$ transitions, a coupling of the structural and electronic states seems plausible, and should be considered as a potential key factor in the charge dynamics of $\theta$-CsZn and $\theta$-CsCo salts.

For the charge-glass state in $\theta$-RbZn, besides the observed close relation between charge crystallization and vitrification there have been no reports on the observation of anomalies related to an charge-glass \textit{transition}. As shown in the Appendix in Fig. \ref{Relative_length_change2}, we have not been able to achieve a quenched charge-glass state in the present thermal expansion study on $\theta$-RbZn. Thus, further investigations are required to clarify whether an anomaly other than the structural glass transitions of the EEGs can be observed in thermal expansion measurements in the presence of a charge-glass state. Such an experiment could possibly be performed on x ray-irradiated samples in which the critical cooling rate $q_{\rm c}^{\rm crit}$ is suppressed and be would help to clarify the relationship between the electron glass and the structural glass of the EEGs in this material.

\section{Summary}
Thermal expansion measurements on $\theta$-(ET)$_2$RbZn(SCN)$_4$ reveal a pronounced first-order phase transition anomaly at the combined charge-order/structural transition at $T_{\rm CO} =215$\,K. In addition, clear evidence for two separate glasslike transitions are observed at $T_{\mathrm{g}}$ = 90--100\,K and $T_{\mathrm{g}}^\dag$ = 120--130\,K. The glass transitions are strikingly similar both in their forms/shapes and temperatures to those reported previously for $\theta$-(ET)$_2$CsZn(SCN)$_4$ and $\theta$-(ET)$_2$CsCo(SCN)$_4$, where the transitions at $T_{\mathrm{g}}$ and $T_{\mathrm{g}}^\dag$ were assigned to a glasslike freezing of the ethylene endgroups of the ET molecules. Our findings indicate that these glasslike transitions are common features in the $\theta$-(ET)$_2MM^\prime$(SCN)$_4$ salts, irrespective of the presence or absence of charge order. Our model calculations on the glasslike dynamics associated with the flexible ethylene endgroups of the ET molecules for the whole $\theta$-(ET)$_2MM’$(SCN)$_4$ family predict two different conformational glass transitions in these salts, consistent with the experimental observations. We also highlight the freezing of ethylene groups as a potential source of quenched disorder, which may significantly impact the charge dynamics in salts where long-range charge order is avoided.

\section*{ACKNOWLEDGMENTS}
This work was supported by the Deutsche Forschungsgemeinschaft (DFG, German Research Foundation) through TRR 288 - 422213477 (Projects A06 and B02). This work was also supported by Grants-in-Aid for Scientific Research (KAKENHI) from MEXT, Japan (Grants No. JP23H01114, JP22H04459, JP21H01793, No. JP20H05144, No. JP19H01833, and No. JP18KK0375), and Grant-in-Aid for Scientific Research for Transformative Research Areas (A) Condensed Conjugation (Grants No. JP20H05869, No. JP21H05471 and JP23H04015) from Japan Society for the Promotion of Science (JSPS). We acknowledge technical assistance by S. Hartmann. Computations were performed using the Wake Forest University (WFU) High Performance Computing Facility, a centrally managed computational resource available to WFU researchers including faculty, staff, students, and collaborators \cite{WakeHPC}.

\begin{figure}[h]
  \includegraphics[width=8.6cm]{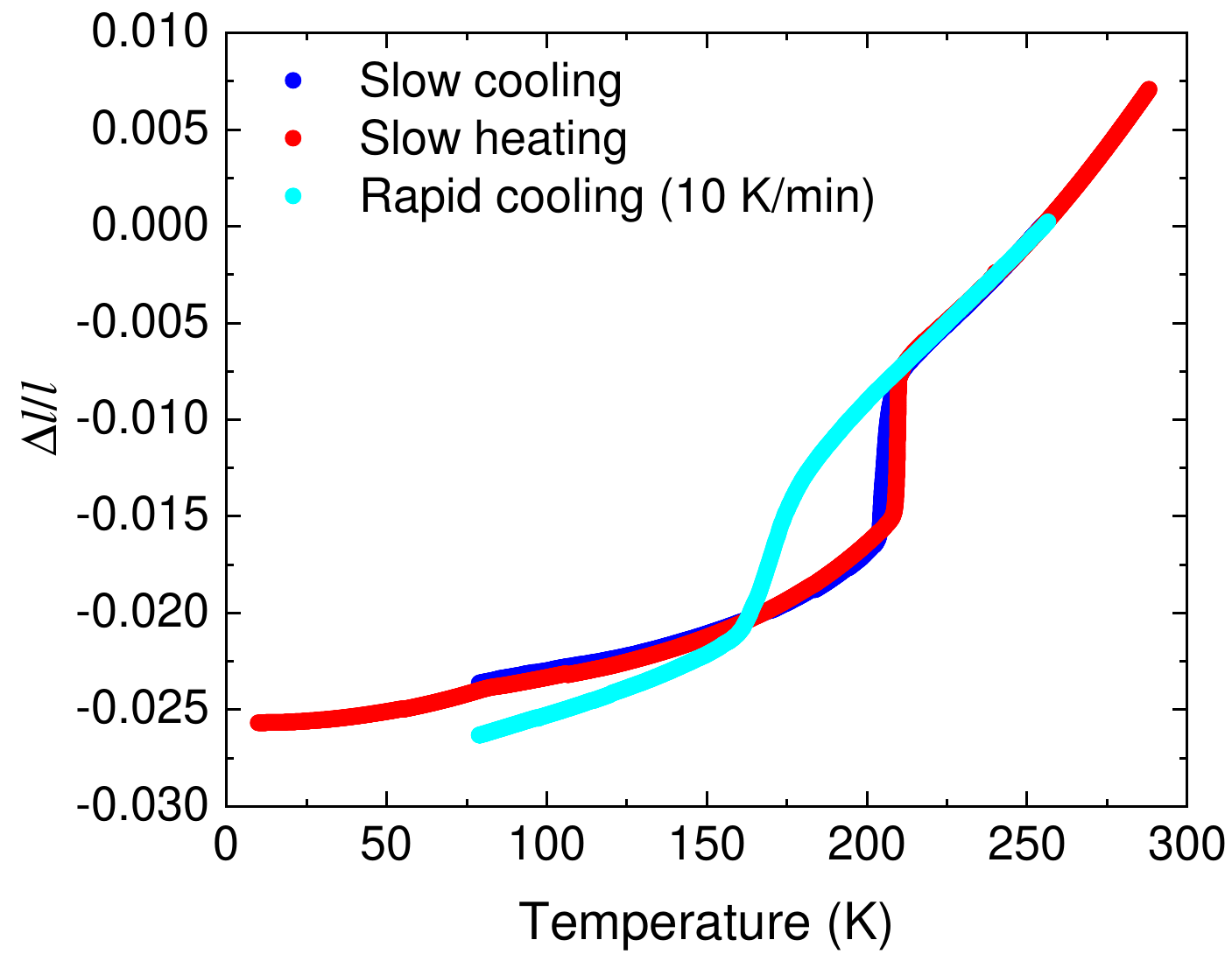}
  \caption{Relative length change for $\theta$-(BEDT-TTF)$_2$RbZn(SCN)$_4$ measured along the $c$ axis for slow cooling (heating) and rapidly cooling, corresponding to a cooling rate of -10\,K/min at 200\,K. }
\label{Relative_length_change2}
\end{figure}
\section*{APPENDIX}
Figure \ref{Relative_length_change2} shows relative length changes of $\theta$-RbZn for slow cooling and rapid cooling, corresponding to a cooling rate of -10\,K/min at 200\,K. At this cooling rate, the discontinuous change in $\Delta l/l$, assigned to the combined charged-order/structural transition, shifts to 170\,K and the transition becomes broader.

\end{document}